\documentclass[amsmath,prl,aps,twocolumn]{revtex4}
\usepackage{amsthm,amsfonts,graphicx,verbatim,epsfig}

\newcommand{\Tr}{{\rm Tr}}

\newcommand{\D}{{\rm d}}

\newcommand{\R}{{\mathord{\mathbb R}}}

\begin{document}
\title{Opposites Attract - A Theorem About The Casimir Force}
\author{Oded Kenneth$^{1}$ and Israel Klich$^{2}$}%
\email{klich@caltech.edu} \affiliation{(1) Department of Physics, Technion, Haifa 32000 Israel \\
(2) Department of Physics, California Institute of
  Technology, MC 114-36 Pasadena, CA 91125}
%\date{February 2002}%
%\author{  }%
%\email{klich@tx.technion.ac.il}
%\affiliation{(1) Department of
%Physics,
%Technion - Israel Institute of Technology, Haifa 32000 Israel \\
%(2) Condensed Matter Department, M.I.T Cambridge, Massachusetts}
%\date{February 2002}%

\begin{abstract}
We consider the Casimir interaction between (non-magnetic) dielectric bodies or
conductors. Our main result is a proof that the Casimir force between two bodies
related by reflection is always attractive, independent of the exact form of the bodies
or dielectric properties. Apart from being a fundamental property of fields, the
theorem and its corollaries also rule out a class of suggestions to obtain repulsive
forces, such as the two hemisphere repulsion suggestion and its relatives.
\end{abstract}

 \maketitle \vskip 2mm
          %%%%%%%%%%%%%%%%%%%%%%%%%%%%%%%%%%%%%%%%
          %               INTRODUCTION           %
          %%%%%%%%%%%%%%%%%%%%%%%%%%%%%%%%%%%%%%%%

%\begin{center}
%
% Department of Physics, \\ Technion - Israel Institute of
%Technology, Haifa 32000 Israel\footnote{e-mail:
%klich@tx.techion.ac.il} \\
%
%
%\end{center}

          %%%%%%%%%%%%%%%%%%%%%%%%%%%%%%%%%%%%%%%%%%%%%%
          %                                            %
          %%%%%%%%%%%%%%%%%%%%%%%%%%%%%%%%%%%%%%%%%%%%%%

%\section{introduction}
The Casimir effect has been a fundamental issue in quantum physics since its prediction
\cite{Casimir48}. The effect has become increasingly approachable in recent years with
the achievement of precise experimental measurements of the effect
\cite{Sparnaay59,Lamoreaux97,MohideenRoy98,Bressi}, probing the detailed dependence of
the force on the properties of the materials, and measuring new variants such as
corrugation effects. The theory and experiment have good agreement for simple
geometries.

In spite of the vast body of work on the subject (For a review see
\cite{BordagMohideenMostepanenko}), some properties of the force are yet under
controversy. Due to the computational complexity of the problem, the main body of work
on the effect is a collection of explicit calculations for simple geometries. In this
Letter we resolve one of these controversies and supply general statements about
Casimir forces, applicable to a broad class of geometries.

The interest in repulsive Casimir and Van Der Waals forces has grown substantially
recently due to possible practical importance in nano science, where such forces may
play a role as a solution to stiction problems. It is known that repulsive forces are
possible between molecules immersed in a medium whose properties are
intermediate between the properties of two polarizable molecules \cite{Isra}.
Conditions for repulsion between paramagnetic materials and dielectrics without
recourse for an intermediate medium were given in \cite{KennethKlich}. However, the
prospect of realizing materials with nontrivial permeability on a large enough
frequency range is unclear \cite{IannuzziAndReply}.

It is common knowledge, based on the Casimir-Polder interaction, that small dielectric
bodies interacting at large distance attract \cite{kn}. Based on summation of two body
forces one may speculate that any two dielectrics would attract at all distances. In
this Letter we show that at least for the case of a symmetric configuration of two
dielectrics or conductors this prediction holds independently of their distance and
shape, for models which can be described by a local dielectric function. Of course, in
any real material as distances become small enough, i.e. compared with interatomic
distances, Casimir treatment of the problem is not adequate anymore.

We first emphasize that the two-body picture is not enough to prove this. Calculations
of the interaction between macroscopic bodies by summation of pair - interactions are
only justified within second order perturbation theory. Indeed, in \cite{KennethKlich}
it was demonstrated how summing two body forces may give wrong prediction for the
sign of interaction between extended bodies.

\begin{figure}
\includegraphics[scale=0.7]{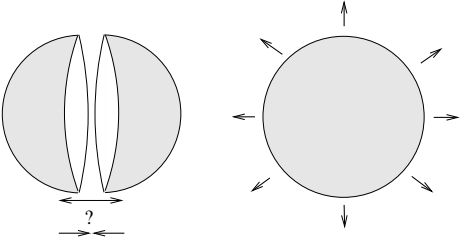}\caption{What is the direction of the force between two conducting hemispheres? While the outward pressure on a
conducting shell might suggest repulsion, it follows from the arguments below that the
hemispheres in fact attract} \label{Hemispheres}
\end{figure}

Another objection to the pair-wise intuition is based on the example of Casimir energy
of a perfectly conducting and perfectly thin sphere. This was worked out by Boyer
\cite{Boyer} and yields an outward pressure on the sphere. This result motivated a
class of suggestions for repulsive forces, the most well known of which are two
conducting hemispheres - considered as a sphere split into two and therefore expected
to repel each other \cite{Lamoreaux97,ElizaldeRomeo} (Fig. \ref{Hemispheres}).

One may try to use perturbative series, such as the multiple scattering series in the
conducting case \cite{BalianDuplantier78} and show the attraction term by term.
However, checking such a claim at orders higher then second might prove a difficult
task. Such an approach is justified for distant bodies, but doesn't seem to be
particularly promising for the problem at hand.

Our main result is that the electromagnetic field (EM) or a scalar field, interacting
with (non-magnetic) bodies, which are mirror image of each other and separated by a
finite distance, will cause the bodies to attract. In particular, this shows that two
hemispheres attract each other. The result holds for a scalar field in any dimension
and even when the bodies are inside an infinite cylinder of arbitrary cross-section
(perpendicular to the reflection plane) with arbitrary boundary conditions (b.c.) on
the cylinder, thereby verifying and generalizing recent results for a Casimir piston
\cite{HertzbergJaffeKardarScardicchio}.

{\bf Expressing the Casimir interaction as a (regular) determinant.} Several
expressions are available for Casimir forces between dielectrics. We find the path
integral method \cite{LiKardar,Kenneth99,FeinbergMannRevzen} a convenient starting
point for the presentation (alternatively, the result may be obtained using other
approaches such as the Green's function method). We start with the case of a scalar
field for simplicity, and explain later how the result is extended to the EM field.
The action of a real massless scalar field in the presence of dielectrics can be
written as
\begin{multline}\label{scalar action}
S={1\over 2}\int\D^{d} {\bf r}\int {\D\omega\over
2\pi}\phi_{\omega}^*(\nabla^2+\omega^2\epsilon({\bf x},\omega))\phi_{\omega}
\end{multline}
where $\phi_\omega^*=\phi_{-\omega}$, and $\epsilon(\omega,{\bf x})=1+\chi({\bf
x},\omega)$ is the dielectric function (we use units $\hbar=c=1$). The change in energy
due to introduction of $\chi$ in the system is formally:
%\begin{multline}\label{formal F}
%F=i\int_0^\infty {\D\omega\over 2\pi}\log\det(\nabla^2+\omega^2+\omega^2\chi(\omega,x))
%\end{multline}
\begin{multline*}\label{formal F}
E_C=E_{\chi}-E_{\chi=0}=\\  -i\int_0^{\infty} {\D\omega\over
2\pi}\log{\det}_{\Lambda}(1+\omega^2\chi({\bf x},\omega) (\nabla^2+\omega^2+i0)^{-1})
\end{multline*}
A determinant is mathematically well defined if it has the form $\det(1+A)$, where $A$
is a "trace class" operator, i.e. $\sum_i |\lambda_i|<\infty$ with $\lambda_i$
eigenvalues of $A$ (For properties see \cite{Simon79}). The expression above is not of
this form, and only has meaning when specifying cutoffs. Removing physical cutoffs will
leave us with an ill defined determinant and so we keep in mind cutoffs at high momenta
in the notation $\det_{\Lambda}$ (one may use instead lattice regularization).

At high frequencies $\chi(\omega,{\bf x})\rightarrow 0$, provides a physical frequency
cutoff. $\chi(\omega)$ and $(\nabla^2+\omega^2+i0)$ are analytic for ${\rm Re}\omega,
{\rm Im} \omega>0$, justifying Wick-rotation of the integration to the imaginary axis
$i\omega$ ending up with:
\begin{multline}
E_{C}=\int_0^{\infty}{\D\omega\over 2\pi}\log{\det}_{\Lambda}(1+\omega^2\chi({\bf
x},i\omega) G_0({\bf x},{\bf x}'))
\end{multline}
Where ${G_0}({\bf x},{\bf x}')=\langle {\bf x}|{1\over -\nabla^2+\omega^2}|{\bf
x}'\rangle$. Restricting the operator $(1+\omega^2\chi G_0)$ to the support of $\chi$
(more precisely to $L^2(Supp(\chi))$) clearly does not affect its determinant. We
assume $\chi$ is nonzero only inside the volumes of the two dielectrics $A,B$ and we
therefore consider in the following $(1+\omega^2\chi G_0)$ as an operator on
${H_A\oplus H_B}\rightarrow{H_A\oplus H_B}$ where $H_A=L^2(A)$ and $H_B=L^2(B)$. It is
then convenient to write it as $(1+\omega^2\chi G_0)\Big|_{H_A\oplus H_B}=\left(
\begin{array}{cc}
1_A+\omega^2\chi_A {G_0}_{AA} & \omega^2\chi_A {G_0}_{AB} \\
\omega^2\chi_B {G_0}_{BA}& 1_B+\omega^2\chi_B {G_0}_{BB} \\
\end{array}
\right)$.
 It turns out that the part of the energy that depends
 on mutual position of the bodies, and as such is responsible for the force, is a
well defined quantity, independent of the cutoffs. To see this, we subtract
contributions which do not depend on relative positions of the bodies $A,B$:
\begin{multline}\label{subtractions}
E_{C}=E_C(A\bigcup B)-E_C(A)-E_C(B)
\end{multline}
As in \cite{FeinbergMannRevzen} this amounts to subtracting the diagonal contributions
to the determinant which are not sensitive to the distance between the bodies, (i.e.
only contributes to their self energies). This yields:
\begin{multline}\label{manipulations for two bodies}
\hspace{-0.5cm} E_{C}= \int_0^{\infty}{\D \omega\over 2\pi} \Big\{\log{\det}_{\Lambda}
\left(\begin{array}{cc}
1+\omega^2\chi_A {G_0}_{AA} & \omega^2\chi_A {G_0}_{AB} \\
\omega^2\chi_B {G_0}_{BA}& 1+\omega^2\chi_B {G_0}_{BB} \\
\end{array}\right)
\\
-\log{\det}_{\Lambda}
\left(\begin{array}{cc} 1+\omega^2\chi_A {G_0}_{AA} & 0 \\
0& 1+\omega^2\chi_B {G_0}_{BB} \\ \end{array}\right)
\Big\}\\
=\int_0^{\infty}{\D \omega\over 2\pi} (\log{\det}_{\Lambda}
\left(\begin{array}{cc} 1 & T_A{G_0}_{AB}\\
T_B{G_0}_{BA}& 1 \\  \end{array}\right)
\end{multline}
where $T_{\alpha}={\omega^2\over 1+\omega^2\chi_{\alpha} {G_0}_{{\alpha}{\alpha}}
}\chi_{\alpha}$.
Finally, using the relation $\det\left(%
\begin{array}{cc}
  1 & X \\
  Y & 1 \\
\end{array}%
\right)=\det(1-YX)$, which holds for block matrices we have:
\begin{eqnarray}\label{F for bulk}&
E_C(a)=\int_0^{\infty}{\D \omega\over 2\pi} \log\det(1-T_A{G_0}_{AB}T_B{G_0}_{BA}).
\end{eqnarray}
Note that the (hermitian) operators $T_{\alpha}$ are exactly the
$T$ operators appearing in the (Wick rotated) Lippmann-Schwinger
equation \cite{scattering books}. Indeed, one may alternatively
derive Eq. \eqref{F for bulk} within Green's function approach and
using $T$ operators.

In \eqref{F for bulk} we disposed of the cutoff $\Lambda$ as the
expression is well defined in the continuum limit. We recall that
an operator $M$ is called {\it positive (denoted $M>0$)} if
$\langle\psi|M|\psi\rangle>0$ for any $\psi$. Since
$\chi(i\omega,x)>0$ (as follows from general properties of the
dielectric function \cite{LifsitzPitaevskii}), the $T$ operators
are positive and bounded, and $T_A{G_0}_{AB}T_B{G_0}_{BA}$ is a
trace class operator without need for cutoffs for any finite
bodies $A,B$ \footnote{This can be verified noting that
$T_A{G_0}_{AB}T_B{G_0}_{BA}$ has a continuous, non-singular kernel
on the compact body $A$ and using M. Reed and B. Simon, {\it
Methods of modern mathematical physics. III (Academic Press)}
$XI.24$}. In fact, this holds also for nonlocal $\chi$ as long as
$f(x)\rightarrow \int_{A}\chi(i\omega,x,x')f(x')\D x'$ is a
bounded positive operator $H_A\rightarrow H_A$. At this point the
determinant is regularized and rigorously well defined for every
$\omega$ and the integration over $\omega$ is convergent due to
the exponential decay of the kernels ${G_0}_{AB}$ as
$\omega\rightarrow\infty$.

It is worthy to note that (for $\chi>0$) {\it all} eigenvalues $\lambda$ of the
(compact) operator $T_A{G_0}_{AB}T_B{G_0}_{BA}$ appearing in \eqref{F for bulk} satisfy
$1>\lambda\geq 0$
\footnote{Note first that $G_0,\chi\geq0$ (as operators) imply $\sqrt{G_0}\chi\sqrt{G_0}\geq 0$
and so its spectrum is contained in the non-negative real axis.
Writing $\sqrt{G_0}T_\alpha \sqrt{G_0}=1-{1\over 1+\sqrt{G_0}\omega^2\chi_{\alpha}\sqrt{G_0}}$
as an operator on $L^2(\R^3)$ it is then clear that its spectrum lies in [0,1).
But since it is hermitian one concludes also $||\sqrt{G_0}T_\alpha\sqrt{G_0}||<1$
from which it follows $||\sqrt{G_0}T_A G_0T_B\sqrt{G_0}||<1$ and hence $\lambda<1$.
Similarly $\sqrt{G_0}T_\alpha\sqrt{G_0}\geq 0$ imply $\lambda\geq 0$.}.

{\bf The Theorem.}
\begin{figure}
\includegraphics[scale=0.4,bb=10 20 500 240,clip]{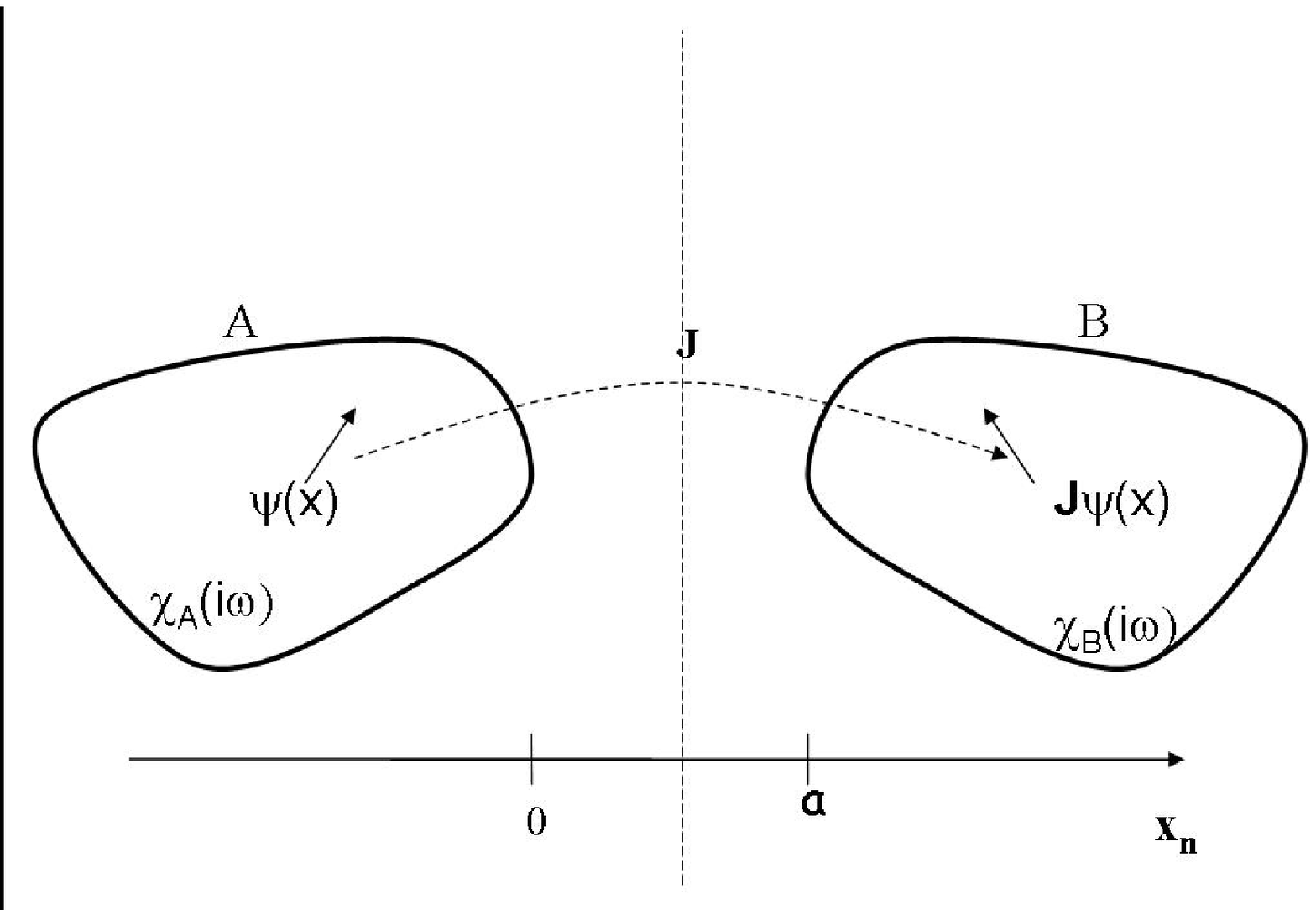}
\caption{Bodies $A$ and $B$ are related by the reflection $J$}\label{shapes}
\end{figure}
Having established a mathematically well defined expression for the Casimir energy, we
now come to the main result: Consider a configuration of two bodies $A,B$ related by a
reflection (Fig. \ref{shapes}), with $\chi(i|\omega|)$ a bounded positive operator and
separated by a finite distance $a$; then (for fixed spatial orientations of the bodies)
$E_C$ given in \eqref{F for bulk} is a monotonically increasing function of $a$ (i.e.
the Casimir force is attractive).

{\it Proof:}
We assume that $A$ is located entirely in the negative $x_n$ half space, and
that $B$ is its mirror image under reflection through the $x_n=a/2$ plane.
To exploit the
reflection symmetry we introduce a mapping $J:A\rightarrow B$ given by $J({\bf
x}_{\perp},x_n)=({\bf x}_{\perp},a-x_n)$. Note that $B=J A$. $J$ is volume preserving
and induces a unitary operator ${\cal J}:H_A\rightarrow H_B$ defined by ${\cal
J}\psi({\bf x})=\psi(J({\bf x}))$. In the case that $\psi$ is a vector field, as in the
EM case below we take ${\cal J}\psi({\bf x})=(\psi_{\perp}(J({\bf x})),-\psi_n(J({\bf
x})))$ (see Fig. \ref{shapes}).
%Note that the distance between $A$ and $B$ is given by
%$a+2min\{|x_n|,x\in A\}$.
Since the bodies $A,B$ are related by reflection we have
$T_{B}={\cal J}T_{A}{\cal J}^{\dag}$ and thus:
\begin{multline}\label{F_C}
E_C=\int_0^{\infty}{\D \omega\over 2\pi} \log\det(1-T_{A}{G_0}_{AB}{\cal J}T_{A}{\cal
J}^{\dag} {G_0}_{BA})
\end{multline}
Note that ${G_0}_{AB}{\cal J}={\cal J}^{\dag}{G_0}_{BA}$ is a hermitian operator (this
can be verified).
The energy can therefore also be expressed as
\begin{multline}\label{YY}
E_C=\int_0^{\infty}{\D \omega\over 2\pi} \log\det
\left(1-(\sqrt{T_{A}}{G_0}_{AB}{\cal J}\sqrt{T_{A}})^2\right)
\end{multline}
We now show by a direct calculation that (as operators on $H_A$):
\begin{eqnarray}& \label{prop1}
{G_0}_{AB}{\cal J}>0  \\ & \label{prop2} \partial_a{G_0}_{AB}{\cal J}<0
\end{eqnarray}
%To show that the Casimir force is attractive, it is sufficient to show that
%$\log\det(1-K^2)$ is a monotonically decreasing function of $a$, irrespective of
%$\omega$. To achieve this we show that the eigenvalues of $K$ are decreasing.
Let $I(a)=\langle\psi|{G_0}_{AB}{\cal J}|\psi\rangle$ for a function $\psi({\bf
x}_{\perp},x_n)\in H_A$. $I(a)$ is explicitly given by
\begin{multline}\label{I}
I(a)= \int_{A\times A} \D {\bf x}\D {\bf x}'\int{\D {\bf k}\over (2\pi)^d} \psi^*({\bf
x})\psi({\bf x}')\\ \times{e^{i {\bf k}_{\perp}\cdot({\bf x}_{\perp}-{\bf
x}_{\perp}')+ik_n(x_n+x_n'-a)}\over {\bf k}^2+\omega^2}
\end{multline}
Note that $x_n+x_n'-a<0$, allowing integration over $k_n$ by closing a contour from below
the real $k_n$ axis:
\begin{multline}\label{I(a)}
I(a)= \int_{A\times A} \D {\bf x}\D {\bf x}'\int{\D {\bf k}_{\perp}\over (2\pi)^{d-1}}
\psi^*({\bf x})\psi({\bf x}')e^{i {\bf k}_{\perp}\cdot({\bf x}_{\perp}-{\bf
x}_{\perp}')}\\ \times{e^{\sqrt{{\bf k}_{\perp}^2+\omega^2}(x_n+x_n'-a)} \over
2\sqrt{{\bf k}_{\perp}^2+\omega^2}}= \int{\D {\bf k}_{\perp}\over
(2\pi)^{d-1}}{e^{-a\sqrt{{\bf k}_{\perp}^2+\omega^2}} \over 2\sqrt{{\bf
k}_{\perp}^2+\omega^2}}
\\
\times\Big|\int_A\D {\bf x} \psi^*({\bf x})e^{i k_{\perp}\cdot {\bf
x}_{\perp}}e^{x_n\sqrt{{\bf k}_{\perp}^2+\omega^2}}\Big|^2
\end{multline}
showing that $I(a)>0$, which proves \eqref{prop1}, and that $\partial_a I(a)<0$ which
%\begin{multline}
%\partial_a I(a)=
%-\int{\D {\bf k}_{\perp}\over 2^d\pi^{d-1}}e^{-a\sqrt{{\bf k}_{\perp}^2+\omega^2}}\\
%\times \Big|\int_A\D x \psi^*({\bf x})e^{i {\bf k}_{\perp}\cdot {\bf
%x}_{\perp}}e^{x_n\sqrt{{\bf k}_{\perp}^2+\omega^2}}\Big|^2<0
%\end{multline}
proves \eqref{prop2}.

%We now come to the last step. Assume bounded hermitian operators
%satisfying $X>0,Y>0$,$\partial_a X=0,\partial_a Y<0$ and $XYXY$ trace class.
%Then $\partial_a\log\det(1-XYXY)>0$
%\footnote{For any eigenvalue of $\lambda_n(a)$ of the positive operator $\sqrt{X}Y\sqrt{X}$
%(with eigenvector $\psi_{n;a}$) , we have by the Feynman-Hellmann theorem that
%$\partial_a\lambda_n(a)=\langle\psi_n|\sqrt{X}(\partial_a Y)\sqrt{X}|\psi_{n}\rangle<0$.
%Since $\sqrt{X}YXY\sqrt{X}$ is trace class, $\log\det(1-XYXY)=
%\log\det(1-(\sqrt{X}Y\sqrt{X})^2)=\sum_n\log(1-\lambda_n^2)$
%is absolutely convergent, and it follows that $\partial_a\sum_n\log(1-\lambda_n^2)>0$.}.
%Taking $X=T_A$ and $Y={G_0}_{AB}{\cal J}$ as in\eqref{F_C} we see that $\partial_a E_C>0$.

From (\ref{prop1},\ref{prop2}) it immediately follows that the
operator $Y=\sqrt{T_A}{G_0}_{AB}{\cal J}\sqrt{T_A}:H_A\rightarrow
H_A$ also satisfy $Y>0,\partial_a Y<0$. Hence a Feynman-Hellman
argument implies that all its eigenvalues $1>\lambda_n(a)\geq 0$
are monotonically decreasing as functions of $a$. Since
$\log\det(1-Y^2)=\sum_n\log(1-\lambda_n^2)$ is absolutely
convergent it follows $\partial_a \log\det(1-Y^2)>0$, and hence by
\eqref{YY} also $\partial_a E_C>0$.

This completes the proof for the scalar case.

To treat the EM case we start with the well known expression Eq.
(80.8) of Lifshitz and Pitaevskii \cite{LifsitzPitaevskii} for the
change in free energy due to variation of the dielectric function
$\epsilon$ at a temperature $T$:
\begin{eqnarray}\label{Lifshitz Pitaevskii expr}\delta F=\delta
F_0+{1\over2}{T}\sum_{n=-\infty}^{\infty} \omega_n^2\Tr ({\cal D}
\delta\epsilon).
\end{eqnarray}

Here $F_0$ is the free energy due to material properties not
related to long wavelength photon field, and $\omega_n=2\pi nT$
are Matsubara frequencies. ${\cal D}$ is the temperature Green's
function of the long wave photon field given by $ {\cal
D}(\vec{x},\vec{x'},i\omega)_{ij}=<\vec{x}|{1\over
\nabla\times\nabla\times+\omega^2\epsilon(r,i|\omega|)}|\vec{x}'>_{ij}
$

Eq\eqref{Lifshitz Pitaevskii expr} may be written as $\delta
F=\delta F_0+\delta F_C$ where \footnote{Alternatively, the EM
case may similarly be derived starting from the functional
determinant corresponding to the EM action. In the axial gauge
${\cal A}_0=0$ this action takes the form:
%\label{vector action}
$S={1\over 2}\int\D^{3} {\bf r}\int {\D\omega\over 2\pi}\vec{{\cal
A}}_{\omega}^*(-\nabla\times\nabla\times +\omega^2\epsilon({\bf
x},\omega))\vec{{\cal A}}_{\omega}.$}
\begin{eqnarray}\label{free energy vector}&
F_C={1\over2}{T}\sum_{n=-\infty}^{\infty}
[\log{\det}_{\Lambda}(\nabla\times\nabla\times+\omega_n^2\epsilon(x,i\omega_n))\\
\nonumber &
-\log{\det}_{\Lambda}(\nabla\times\nabla\times+\omega_n^2)]\\
\nonumber & ={1\over2}{T}\sum_{n=-\infty}^{\infty}
\log{\det}_{\Lambda}
(1+\omega_n^2\chi(x,i\omega_n){\cal D}_0(i\omega_n)).
\end{eqnarray}
where $ {\cal
D}_0(\vec{x},\vec{x'},i\omega_n)_{ij}=<\vec{x}|{1\over
\nabla\times\nabla\times+\omega_n^2}|\vec{x}'>_{ij}$. Note that
$F_C$ is exactly the same as \eqref{formal F}, with the scalar
propagator $G_0$ replaced by the vector propagator ${\cal D}_0$.

%We also assume that $\delta F_0=0$ for the kind of variations we
%make, which do not involve deformation of the bodies.
Thus, starting with this expression, one repeats
\eqref{subtractions} and \eqref{manipulations for two bodies} to
get \eqref{F for bulk}, replacing $G_0$ by ${{\cal D}_0}$
everywhere (including in the definition of the $T$ operators). The
analysis of the determinant now proceeds exactly as in the scalar
case. The only place in the proof which needs to be modified is
where the explicit form of $G_0$ was used i.e. Eq.\eqref{I}, where
we now have to use ${{\cal D}_0}_{ij}(k,i\omega)={1\over
k^2+\omega^2}(\delta_{ij}+{k_ik_j\over\omega^2})$ instead.
%, while
%keeping in mind that now all operators are $3\times 3$ operator
%valued matrices.

The effect of using the vectorial propagator in Eq.\eqref{I} is to replace $\psi^*({\bf
x})\psi({\bf x}')$ by $\psi^*_i({\bf x})\psi_j({\bf x}')
(\delta_{ij}+{k_ik_j\over\omega^2})$. In the vectorial case ${\cal J}$ acts by ${\cal
J}\psi({\bf x})=(\psi_{\perp}(J({\bf x})),-\psi_n(J({\bf x})))$ so we get a factor
$(-1)^{\delta_{jn}}$. Substituting and integrating over $k_n$ as before we find
\begin{multline}I_{vec}(a)=\pi
\int{\D {\bf k}_{\perp}\over (2\pi)^3} {e^{-a\sqrt{{\bf
k}_{\perp}^2+\omega^2}}\over\sqrt{{\bf k}_{\perp}^2+\omega^2}}
\times\\
\left[(-1)^{\delta_{jn}}\phi^*_i\phi_j(\delta_{ij}+{k_ik_j\over\omega^2})\right]
\Big|_{k_n=-i\sqrt{{\bf k}_{\perp}^2+\omega^2}}\end{multline} where $\phi_j({\bf
k}_{\perp})=\int_A\D {\bf x} \psi_j({\bf x})e^{-i {\bf k}_{\perp}\cdot {\bf
x}_{\perp}}e^{x_n\sqrt{{\bf k}_{\perp}^2+\omega^2}}$. Now it is straightforward to
check that the expression in square brackets is positive for any $\phi_i$ and the
theorem follows.

{\bf Extensions and remarks:} \\ 1) {\it Finite temperatures:} As remarked above we
have $\int{\D\omega\over 2\pi}\rightarrow T\sum_{\omega_n}$ at finite $T$.
Since the positivity arguments apply to the determinant at each fixed imaginary
frequency $\omega$, they will also hold at finite $T$.

2) {\it Confined geometry in transverse directions} Our theorem is
easily extended to cover the case when placing the system inside
an infinite cylinder, perpendicular to the $x_n=0$ plane, with
arbitrary cross section. In this case, one has to replace our
$G_0$ by the appropriate Helmholtz green's function in the
cylinder: $G_0(x,x')=\int\D k_n\sum_j {\varphi_j({\bf
x}_{\perp})\varphi_j^{\dag}({\bf x'}_{\perp})\over
\omega^2+k_n^2+E_j}e^{i k_n(x_n-x_n')}$, where $\varphi_n(x)$ are
the appropriate quantized eigenmodes in the transverse direction,
and the integration over ${\bf k}_{\perp}$ is replaced by discrete
summation. Substituting this expression in the relevant integrals
such as \eqref{I(a)} yields the attraction. Since the attraction
is independent of the $\varphi_j$, this result is independent of
the b.c. one sets on the containing cylinder.

3) {\it Dielectric in front of mirror} Suggestions were raised for repulsion between
arrays of dielectrics and a mirror plane \cite{GussoSchmidt04}, based on results for a
rectangular cavity. Variation of our theorem shows that one actually has attraction.
Consider the body $A$ to the left of a Dirichlet mirror located at $x_n=a/2$. By the
image method the propagator is replaced by $G({\bf x},{\bf x'})=G_0({\bf x}-{\bf
x'})-G_0({\bf x}+{\bf x'}-a\hat{n})$.
This may also be written as $G-G_0=-G_0{\cal J}$.
It is then straightforward to arrive at the
expression for the energy\footnote{An alternative way to derive this relation is to
substitute $\chi_B=\lambda\theta(x_n-a/2)$ into eq\eqref{F for bulk} and consider the
limit $\lambda\rightarrow\infty$.}
 analogous to \eqref{F for bulk} with $\det(1- G_0{\cal J}T_A)$
replacing  $\det(1- G_0T_AG_0T_B)$.
Using similar considerations as
in the proof above the attraction follows.

4) {\it Dirichlet b.c.:} Our approach never uses directly b.c. on
the dielectrics; instead, we consider interaction with an
arbitrary permittivity $\epsilon(x,\omega)$. This is adequate for
describing real conductors. Idealized Dirichlet b.c. for a scalar
field and ideal conductor b.c. for EM field are obtained as the
limit of large $\chi(i\omega)$; However, Neuman b.c. do not follow
from the present treatment, since they do not correspond to a
positive perturbation, or indeed to any regular perturbation.

5) {\it Nonpositive perturbations.} Cases of effective $\chi<0$
typically occur when the medium between the bodies has higher
permittivity then the bodies. These cases as well as cases with
nontrivial magnetic permeability $\mu$ may be covered in a way
similar to the above theorem. However conditions on $\chi,\mu$
must be specified to ensure that the eigenvalues of
$1-T_AG_0T_BG_0$ remain positive. These conditions are related to
the assumption that the perturbation may not be so negative as to
introduce negative energy modes into the system.

{\bf Summary:} Our main result is that the Casimir force between
two dielectric objects, related by reflection, is attractive. Our
theorem serves as a no-go statement for a class of suggestions for
repulsive Casimir forces. Of course, the treatment is only valid
at distances where the system may be described reliably in terms
of the field and local dielectric functions alone. Although the
above proof applies only to symmetric configurations, the approach
presented here may be used to analyze the more general cases. A
natural question rises: How far can our result be generalized?
Which classes of interacting fields obey it?

I.K would like to thank L. S. Levitov R. L. Jaffe and A.
Scaricchio for discussions. O.K. is supported by the ISF.

\end{document}